# EXPERIMENTAL SINGLE ELECTRON 4D TRACKING IN IOTA *

A. Romanov[†], J. Santucci, G. Stancari,
Fermi National Accelerator Laboratory, Batavia, IL 60510, USA

*Abstract*

This paper presents the results of the first experiments on 4D tracking of a single electron using a linear multi-anode photomultiplier tube. The reported technology makes it is possible to fully track a single electron in a storage ring, which requires tracking of amplitudes and phases for both, slow synchrotron and fast betatron oscillations. Complete tracking of a point-like object enabled the first direct measurements of single-particle dynamical properties, including dynamical invariants, amplitude-dependent oscillation frequencies, and chaotic behavior.

## INTRODUCTION

Complete tracking of a charged particle in a circular accelerator will enable a new class of diagnostics capabilities. It will allow measurements of important single-particle dynamical properties, including dynamical invariants, amplitude-dependent oscillation frequencies, and chaotic behavior. The true single-particle measurements can be employed for benchmarking of long-term tracking simulations, for training of AI/ML algorithms, and ultimately for precise predictions of dynamics in present and future accelerators.

Observation of a single electron in storage rings has a long history that goes back to experiments at AdA, the first electron-positron collider [1, 2]. Several experiments using various instruments were done in the past to track single electron dynamics in storage rings, with the goal to track relatively slow synchrotron oscillations [3–5] and tracking of all 3 mode amplitudes [6].

The goal of the study presented here was to demonstrate for the first time a complete 6-dimensional tracking of a single particle, an electron in our case, in a storage ring. Unfortunately, due to long delivery times we were able to use only one coordinate sensitive photon detector. This allowed us to track an electron in 4 dimensions of the phase space, covering longitudinal and horizontal planes.

## EXPERIMENTAL SETUP

Each of the 8 main dipoles in IOTA [7] is equipped with synchrotron light stations installed on top of the magnets themselves. The light out of the dipoles is deflected upwards and back to the horizontal plane with two 90-degree mirrors. After the second mirror, the light enters the dark box, which is instrumented with customizable diagnostics, as shown in Figure 1. A focusing achromatic lens with a 40 cm focal length and an iris are installed in the vertical insulation tube that connects to the mirror holders. This experiment used one of such diagnostics stages located at the M3L dipole.

___________________
* This work is supported by the U.S. Department of Energy, Office of Science, Office of Basic Energy Sciences, under Contract No. DE-AC02-06CH11357 and by Fermilab's Laboratory Directed Research and Development grant FNAL-LDRD-2022-041.
† aromanov@fnal.gov

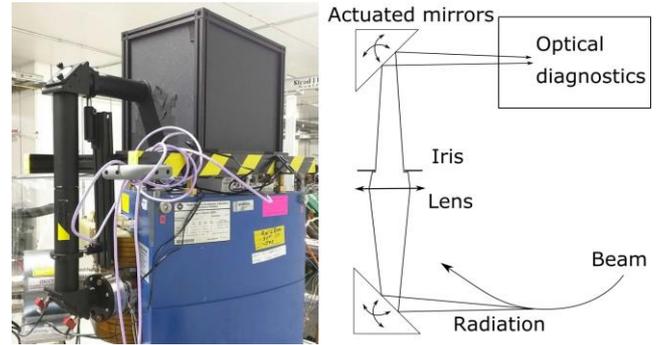

Figure 1: Photograph of the optical diagnostics setup at one of the IOTA's main dipoles (left) and corresponding schematic diagram (right).

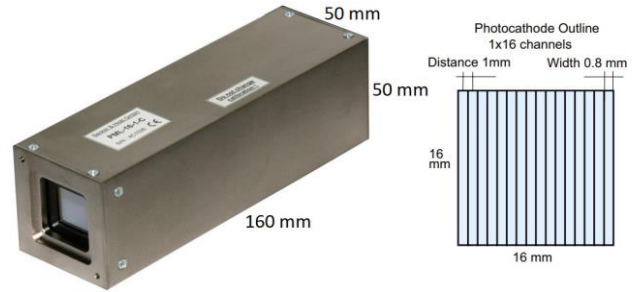

Figure 2: General view of the PML-16 multianode PMT (left) and geometry of its sensitive area (right).

The PML-16 detector from the Becker&Hickl company based on multianode photomultiplier tube (PMT) was used for the presented experiment. Figure 2 shows general view of the detector and its dimensions, including the geometry of the sensitive area. The PML-16 detectors have active area of 16x16 mm with 16 individual cathodes arranged in a linear array. To fully utilize this relatively large area a defocusing lens was added to the optical system with the goal to make the larger beam sigma be around 2 mm when focused on the sensitive area of PMT (either in horizontal or vertical plane).

PML-16 has a preamp and channel encoding electronics attached to the PMT forming a single unit. This allows minimisation of the noise and time jitter. Control over the detector's high voltage is done by the DCC-100 card. The SPC-130 card is measuring time of arrival and position of a segment that detected a photon. Figure 3 shows connection layout of the PML-16 detector.

A modification to the existing optical and mechanical systems was done to match beam and the detector sizes. Figure 4 shows the layout of the instruments. The setup allows keeping existing operational modes, while enabling single electron tracking with two magnification factors, the nominal 88% and 400% necessary to effectively use large aperture of the PML-16 detector.

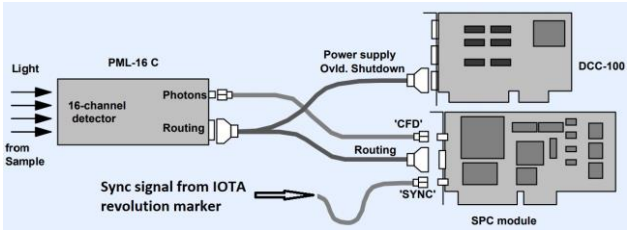

Figure 3: General connection scheme of the PML-16 detector. DCC-100 is a voltage control and overload protection unit that can control two detectors. SPC modules are used to record intra-cycle time, consecutive number of cycle, and a position of the segment for the detected photons. Each detector requires one SPC module.

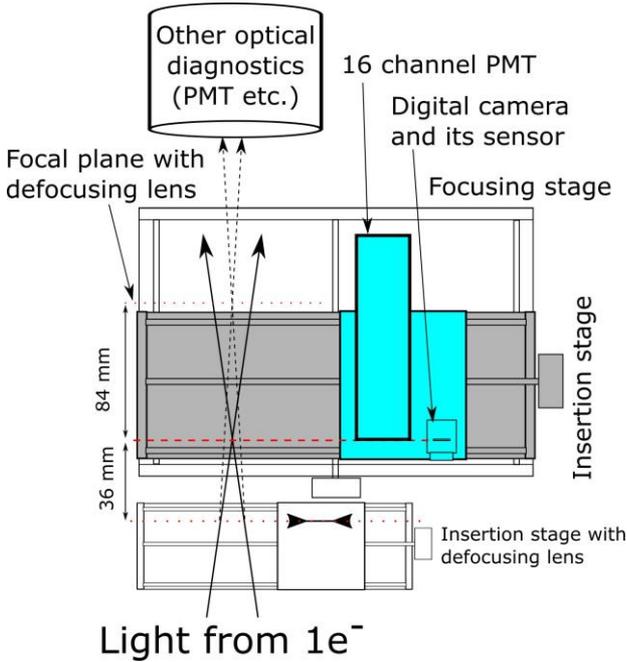

Figure 4: Schematic diagram of the opto-mechanical setup for an electron tracking. Both 16ch PMT and a digital camera (blue shaded) are located on a stack of movable stages. Focusing stage can move insertion stage (grey shaded) to position sensors in the focal planes. The insertion stage can position either one of the sensors on axis of the light beam or let the light pass through to other detectors. Additional insertion stage can move a defocusing lens in and out of the photons path changing magnification factor from 88% to 400% which matches beam size to the size of 16ch PMT.

The reported measurements were done concurrently with other experiments that were ongoing in the IOTA ring without any special modifications to the lattice parameters. The only implemented modification was small change of the horizontal betatron tune to move the working point off the coupling resonance and have a flat beam. Resulting IOTA parameters are listed in TABLE. The lattice was well characterized using LOCO method. The following tolerances are expected for the points of observation of the optical instruments:

- Beta functions accuracy of 5%
- Dispersion functions error smaller than 1 cm
- Betatron tunes within 0.001

Table 1: IOTA parameters during the experiment.

| Parameter | Value |
|---|---|
| Perimeter | 39.96 m |
| Momentum | 150 MeV/c |
| Bunch intensity | 1 $e^-$ |
| RF frequency | 30 MHz |
| RF voltage | 350 V |
| Betatron tunes, ($\nu_x, \nu_y$) | (5.2965, 5.3) |
| Synchrotron tune, $\nu_s$ | $3.5 \times 10^{-4}$ |
| Damping times, ($\tau_x, \tau_y, \tau_s$) | (2.08, 0.65, 0.24) s |
| Horizontal emittance, $\epsilon_x$ | 127 nm |
| Momentum spread, $\Delta p/p$, RMS | $1.3 \times 10^{-4}$ |
| Momentum compaction, $\alpha_p$ | 0.083 |
| Natural chromaticity $C_x, C_y$ | -10.9, -9.4 |

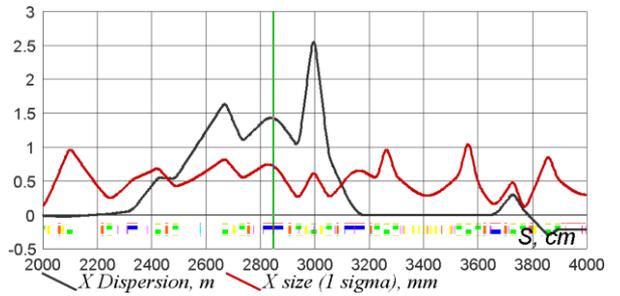

Figure 5: Horizontal beam size and horizontal dispersion of the IOTA lattice. The second half of the ring is shown, ending at the injection straight section. The vertical green line shows location of the M3L monitor.

Figure 5 shows horizontal beam size and dispersion for the used IOTA configuration.

## RESULTS

The data set used for the presented results consists of a set of 3 numbers for each of the detected photons: the turn number at which the detection happened, the intra-turn time and the number of a segment that detected the photon. A total of 102767 photons related to the electron have been detected over 10 second time window.

Figure 6 shows an example of tracking an electron in 4D phase space over about 60000 turns using 80 photons detected over that time. Table 2 contains corresponding trajectory parameters assuming harmonic oscillations in both longitudinal and horizontal directions. Uncertainties of the trajectory parameters were calculated using bootstrap method.

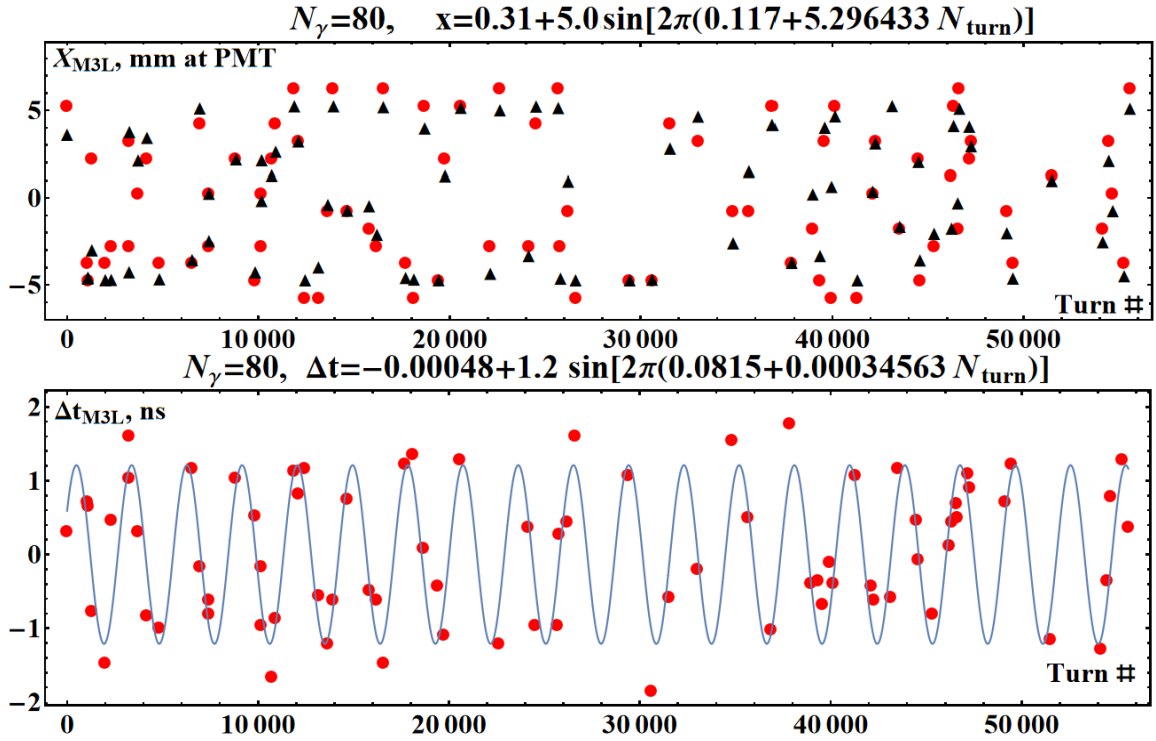

Figure 6: (Top) Horizontal positions of an electron measured (red circles) and reconstructed on the same turns (black triangles) assuming harmonic oscillations. (Bottom) Reconstruction of the synchrotron oscillations (solid line) compared to the measured delays of the arrival time (red circles). Same photon detection events were used for both plots.

Table 2: Parameters of the 4D electron trajectory measured using 80 photons over 60000 turns.

| Parameter | Value |
| --- | --- |
| Horizontal betatron tune | 5.2964325(5) |
| Horizontal betatron phase | $2\pi \cdot 0.12(2)$ |
| Horizontal betatron amplitude | 5.0(2) mm |
| Synchrotron tune | 0.0003456(7) |
| Synchrotron phase | $2\pi \cdot 0.09(2)$ |
| Synchrotron amputude | 1.20(8) ns |

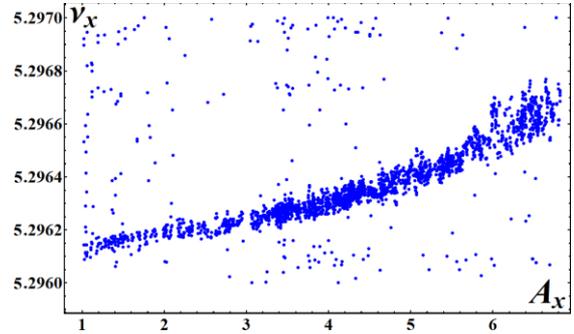

Figure 7: Dependence of the betatron tune on the amplitude of the horizontal oscillations at the image plane.

Because of random kickbacks from the synchrotron radiation photons oscillation amplitude change in time which allows naturally scan the phase space. Figure 7 shows dependence of the horizontal betatron tune on the horizontal betatron amplitude from the analyzed data set. The Fourier transform was used to extract spectrum from 20 photons with a fine peak detection in the range between tunes 5.296 and 5.297. This simplified algorithm increased speed of the analysis but resulted in a higher noise in the reconstructed parameters. Another complication for the analysis is that we have a jitter coming from power supplies that varies betatron tunes and has been filtered out for the presented amplitude dependence plot.

## SUMMARY

Presented results are the first experimental tracking of betatron and synchrotron oscillations at the same time for a single electron in a storage ring. This 4D tracking proves that with addition of the second coordinate sensitive single photon detector it will be possible to fully track an electron in a storage ring.

As an example of a practical use, betatron tune was measured with exceptional precision of $5 \ast 10^{-7}$ as well as a dependence of horizontal betatron tune on the oscillations amplitude.